\documentclass[aps,pra,showpacs,preprintnumbers,amsmath,amssymb,superscriptaddress,floatfix]{revtex4}
\usepackage{graphicx}
\usepackage{dcolumn}
\usepackage{bm}
\begin{document}
\renewcommand{\thefigure}{\arabic{figure}}
\title{Two-dimensional boson-fermion mixtures}
\author{A. L. Suba{\c s}{\i}}
\affiliation{Department of Physics, Bilkent University,
Bilkent, Ankara 06800, Turkey}
\author{S. Sevin{\c c}li}
\altaffiliation[Current address: ]{
Max-Planck-Institut f{\"u}r Physik Komplexer
Systeme, N{\"o}thnitzer Str. 38, 01187 Dresden, Germany}
\affiliation{Department of Physics, Bilkent University,
Bilkent, Ankara 06800, Turkey}
\author{P. Vignolo}
\affiliation{Institut Non Lin{\'e}aire de Nice, Universit{\'e}
de Nice-Sophia Antipolis, CNRS, 1361 route de Lucioles, 06560
Valbonne, France}
\author{B. Tanatar}
\affiliation{Department of Physics, Bilkent University,
Bilkent, Ankara 06800, Turkey}

\begin{abstract}
Using mean-field theory, we study the equilibrium properties of 
boson-fermion mixtures confined in a harmonic pancake-shaped trap 
at zero temperature.  When the modulus of the $s$-wave scattering 
lengths are comparable to the mixture thickness, two-dimensional 
scattering events introduce a logarithmic dependence on density 
in the coupling constants, greatly modifying the density profiles 
themselves. We show that for the case
of a negative boson-fermion three-dimensional $s$-wave scattering 
length, the dimensional crossover stabilizes the mixture against 
collapse and drives it towards spatial demixing.
\end{abstract}
\maketitle

\section{Introduction}
Fermionic atomic gases were brought together with bosonic atoms 
to quantum degeneracy in several alkali atom mixtures, such 
as $^7$Li-$^6$Li \cite{Truscott2001a,Schreck2001a}, 
$^{23}$Na-$^6$Li \cite{Hadzibabic2002a}, $^{87}$Rb-$^{40}$K 
\cite{Goldwin2002a,Roati2002a,Modugno2002a}, and very recently
in a mixed gas of ytterbium (Yb) isotopes, $^{174}$Yb-$^{173}$Yb
\cite{Fukuhara2009}.
The boson-fermion (BF) coupling strongly affects the equilibrium 
properties of the mixture and can drive quantum phase transitions, as 
collapse \cite{Modugno2002a} in the presence of attractive BF 
interaction, or spatial demixing as recently
observed in the context of three-dimensional (3D)
atomic fermion - molecular boson mixtures 
\cite{Partridge2006a,Shin2006a}, where the 
strong interspecies repulsion leads to phase separation. 

Such mixtures can be realized from an imbalanced two-component 
Fermi gas ($^{40}$K-$^{40}$K or $^6$Li-$^6$Li mixtures) where all
minority fermions become bound as bosons and form a Bose-Einstein 
condensate (BEC). Though imbalanced Fermi gases allowed to 
observe spatial phase separation between bosonic dimers and fermions, 
the advantage of a two atomic species
BF mixture is that boson-boson (BB) and BF interactions
can be driven independently and that one can access attractive BF 
interactions \cite{Ospelkaus2006a,Zaccanti2006a}.

The structure and the stability of trapped BF mixtures were studied
in 3D by using the Thomas-Fermi approximation for the bosonic
component \cite{Molmer1998,Akdeniz2002} and by using a modified
Gross-Pitaevskii (GP) equation for the bosons which self-consistently
includes the mean-field interaction generated by the fermionic cloud
\cite{Roth2002,pelster}. Effects of the geometry induced by the 
trap deformation
were studied in the Thomas-Fermi regime in a quasi-3D limit, i.e.
when collisions can be still considered as 
three-dimensional \cite{Akdeniz2004}. Such a simple
model predicts, in a pancake-shaped trap, that the stability of 
the mixture depends only on the scattering length 
and the transverse width of the cloud. One should expect, in a true 
dimensional crossover, namely including dimensional effects in 
scattering events, that the mixture stability depends critically 
on the energy, and thus on the number of particles.

The dimensional crossover from a 3D to a 2D trapped mixture may 
be studied in the experiments by
flattening magnetic or dipolar confinements \cite{Gorlitz2001}, 
or by trapping atoms in specially designed pancake potentials, 
as rotating traps \cite{Schweikhard2004},
gravito-optical surface traps \cite{Rychtarik2004}, 
rf-induced two-dimensional traps \cite{Colombe2004}
or in one-dimensional lattices \cite{Stock2005} where a 3D gas can be
split in several independent disks.

In the limit where scattering events are bi-dimensional, it is 
well known that a hard-core boson gas shows very different features 
from its 3D counterpart. In 3D, particle interactions can be 
described by the zero-momentum and zero-energy limit of $T$-matrix, 
leading to a constant coupling parameter. In 2D, $T$-matrix vanishes 
at low momentum and energy \cite{Schick1971,Popov1966} and the 
first-order contribution to the coupling is obtained by taking 
into account the many-body shift 
in the effective collision energy of two-condensate atoms 
\cite{alkhawaja,lee2}. 
This leads to an energy dependent coupling parameter that
greatly affects the equilibrium and the dynamical properties of the 
gas \cite{Tanatar2002,Hosten2003}.

In this paper we study the equilibrium properties of a mixture of 
condensed bosons and spin-polarized fermions, through the dimensional 
crossover from three to two dimensions, by following the procedure 
outlined by Roth \cite{Roth2002} for the 3D mixture. We neglect 
fermion-fermion interactions and we include BF $s$-wave 
interaction self-consistently in a suitably modified 
GP equation for the bosons. For the case of BF 
repulsive interaction, the increasing anisotropy softens the 
repulsion, and a quasi-3D spatially demixed mixture is mixed in 
quasi-2D. For the case of BF attractive 
interactions, the dimensional crossover acts as a Feshbach resonance
and induce repulsive interactions, so that a Q3D mixture near collapse
can be driven towards spatial demixing in Q2D.
In the strictly 2D regime the results depend on the model one assumes 
for the bi-dimensional scattering lengths.

The paper is organized as follows. In Sec. \ref{model} we 
introduce the theoretical mean-field model for the description of 
ground-state density profiles
of the BF mixture. The models for the coupling through the dimensional 
crossover are outlined in Sec. \ref{interaction}. The density 
profiles obtained for a $^6$Li-$^7$Li and a $^{40}$K-$^{87}$Rb 
mixtures are shown in Sec. \ref{results}. Section \ref{summary} 
offers a summary and some concluding remarks.

\section{Mean-field model for the density profiles}\label{model}
We consider a BF mixture in a 2D geometry, with respective
particle numbers $N_B$ and $N_F$, confined in harmonic trap potentials
$V_{B,F}=\frac{1}{2}m_{B,F}\omega_{B,F}^2r^2$. Here $m_{B,F}$
is boson (fermion) mass and $\omega_{B,F}$ is the radial trap frequency  as
seen by boson or fermion species. Within the mean-field approach
the total energy functional at $T=0$ is written as
\begin{equation}
\begin{split}
E[\psi_B,\psi_F]&=\int\,d^2r \left\{
\frac{\hbar^2}{2m_B}|\nabla \psi_B|^2+V_B(r)|\psi_B|^2+
\frac{1}{2}g_{BB}|\psi_B|^4\right\} \\
&\quad +\int\,d^2r \left\{T_F+V_F(r)|\psi_F|^2\right\}
+\int\,d^2r\, g_{BF}|\psi_B|^2|\psi_F|^2,
\end{split}
\end{equation}
where $\psi_{B,F}$ is the ground-state wave function of bosons
and fermions, respectively. In the above boson species are in
the condensed state and fermion species is 
assumed to be spin-polarized and its kinetic energy is written
within the Thomas-Fermi-Weizsacker approximation as
\cite{zaremba,jezek}
\begin{equation}
T_F=\frac{\hbar^2}{m_F}\left(\pi n_F^2+
\frac{\lambda_W}{8}\frac{|\nabla n_F|^2}{n_F}\right)\, ,
\end{equation}
where $n_F=|\psi_F|^2$ is the fermion density and the Weizsacker
constant is $\lambda_W=1/4$. 
Normalization conditions for $N_B$ bosons and $N_F$ fermions read 
$N_B=\int d^2r|\psi_B|^2$ and $N_F=\int d^2r|\psi_F|^2$. The interaction
couplings between the bosons and between bosons and fermions are denoted by
$g_{BB}$ and $g_{BF}$, respectively.
One notable difference between the form of the energy
functional given above and that in 3D, is that the BB and 
BF interaction strengths are in general density dependent in
contrast to the situation in 3D. More specifically, in 3D the 
interaction strengths are proportional to the scattering
lengths $a_{BB}$ and $a_{BF}$ whereas in 2D as we shall explain
below they depend on the density or equivalently the chemical potential.
The Euler-Lagrange equations for the mixture read \cite{jezek}
\begin{equation}
\left\{-\frac{\hbar^2}{2m_B}\nabla^2+V_B+g_{BB}|\psi_B|^2+g_{BF}|\psi_F|^2
-\mu_B\right\}\psi_B=0\, ,
\label{dens1}
\end{equation}
and
\begin{equation}
\left\{-\frac{\hbar^2}{2m_F}\lambda_W\nabla^2+V_F+\frac{\hbar^2}{m_F}2\pi|\psi_F|^2
+g_{BF}|\psi_B|^2-\mu_F\right\}\psi_F=0\, ,
\label{dens2}
\end{equation} 
in which we have introduced the chemical potentials $\mu_{B,F}$ 
for bosons and fermions. The above equations of motion are obtained 
by functional differentiation from $E[\psi_B,\psi_F]$ neglecting the
higher-order terms involving $\delta g/\delta\psi_{B,F}$ which is
valid in the dilute gas limit $n_Ba_{BB}^2\ll 1$ and
$n_Fa_{BF}^2\ll 1$. The dilute gas conditions above further
maintain that beyond mean-field corrections are not called for.
They can become notable when $a_{BB},\, a_{BF}$ and/or
$N_B,\, N_F$ are large for fixed trap frequencies. For the
systems under consideration we have chosen the parameters
appropriately and verified by numerical calculations so that
$n_Ba_{BB}^2,\,n_Fa_{BF}^2\ll 1$. Therefore, in the 
examples we shall discuss subsequently, the beyond mean-field
terms in the energy functional are not important.

It should also be noted that the existence of BEC in 2D needs to be 
treated carefully. Initial attempts have concluded that no BEC
could occur in 2D trapped gases but
recent considerations within the Hartree-Fock-Bogoliubov 
approximation, \cite{gies_PRA}
the density dependent interaction strength \cite{bhaduri} and numerical 
simulations \cite{markus}
have established firmly the occurrence of BEC for such systems.
Thus, our assumption of a 2D condensate at $T=0$ is justified.

\section{2D Interaction Models}\label{interaction}
In cold atom experiments a 2D geometry is obtained by trapping the atoms
in a highly anisotropic trap where the axial confinement is very tight,
so that the axial potential is on the same order or
larger than the chemical potentials
of the two components. Within this condition, the axial widths are 
on the order of the oscillator lengths for the axial direction
$a_{jz}=\sqrt{\hbar/m_j\omega_{jz}}$, $\omega_{jz}$ being the axial trap 
frequency for bosons ($j=B$) and for fermions 
($j=F$). For simplicity, here and in the following we assume that 
$a_{Bz}=a_{Fz}=a_z$. 

The value of $a_z$ with respect to the modulus of the 3D scattering 
lengths, determines whether the scattering events occur in 3D or in 2D, 
and thus suggests how to calculate the many-body interaction potentials. 

The interaction couplings $g_{BB}$ and $g_{BF}$ are determined
microscopically from the effective interaction potentials (two-body
scattering amplitude, $T$-matrix, etc.) in the limit of
low energy and momenta. In the case of a 3D system, the scattering
amplitude and $g_{BB}$ and $g_{BF}$ are constants determined by the
$s$-wave scattering lengths $a_{BB}$ and $a_{BF}$. In 2D the scattering
theory approaches give rise to a logarithmic dependence
\cite{Schick1971,Popov1966}. Starting from a 3D system and increasing the
anisotropy (by increasing the trap frequencies in the axial direction)
the geometry flattens to take a pancake shape and eventually a genuine
2D system is 
obtained. In the following we identify different scattering regimes
depending on the relation between the axial confinement length and 
scattering lengths and provide expressions for the interaction 
couplings in these regimes.

\subsection{Quasi-3D scattering}
In this regime, the axial oscillator length $a_z$
of the mixture is assumed to be larger than the modulus
of $a_{BB}$ and $a_{BF}$, the $s$-wave scattering 
lengths for BB and BF interactions, respectively. 

The effective BB interaction strength can be obtained by 
multiplying the 3D value of the coupling
$g_{BB}^{3D}=4\pi\hbar^2 a / m_B$
with  a factor $|\phi(0)|^2=1/\sqrt{2\pi} a_z$, $\phi(z)$ being
the axial wavefunction. This is obtained by
assuming that the motion in the $z$-direction is frozen in the ground
state of the harmonic potential with trapping frequency $\omega_{jz}$ and
integrating the 3D GP equation over $z$ (after multiplying with
$\phi^*(z)$ in the spirit of taking an expectation value. The chemical
potential $\mu$ gets shifted by $\hbar\omega_{jz}/2$).
Assuming that the profile for fermions to also be Gaussian 
in the $z$-direction, we apply the same idea to
the BF interaction $g_{BF}^{3D}=2\pi\hbar^2 a / m_{BF}$, 
where $m_{BF}$ is the reduced mass. Thus, we obtain
\begin{equation}
g_{BB}=\frac{2\sqrt{2\pi}\hbar^2}{m_B}\frac{a_{BB}}{a_z}
\, ,\,\,\,\hbox{and}\,\,\,
g_{BF}=\frac{\sqrt{2\pi}\hbar^2}{m_{BF}}\frac{a_{BF}}{a_z},
\label{int_3D}
\end{equation}
as the effective interaction couplings in the quasi-3D
scattering regime.
 
\subsection{Strictly-2D scattering}
This regime corresponds to the limit $a_z\ll |a_{BB}|,|a_{BF}|$.
The coupling parameter we use is from a $T$-matrix
calculation \cite{alkhawaja,lee2,lee} which takes into account the 
many-body shift in
the effective collision energy of two condensate atoms and it becomes a
self-consistent problem.
Since $a_z< |a_{BB}|$, $|a_{BF}|$ the calculation is purely 2D, 
and the interaction strengths do not depend on the 
parameters in the $z$-direction. Al Khawaja {\it et al}.
\cite{alkhawaja} argue that when two condensate atoms collide at
zero momentum they both require an energy $\mu_B$ to be excited
from the condensate and thus the many-body coupling is given
by evaluating at
$-2\mu_B$ the two-body $T$-matrix ($T_{2b}$) setting
$g_{BB}=\langle 0|T_{2b}(-2\mu_B)|0\rangle$. On
the other hand, Lee {\it et al}. \cite{lee} calculate
$T_{2b}$ at $-\mu_B$ arguing that this result includes the
effect of quasiparticle energy spectrum of the intermediate
states in the collision. Gies {\it et al}. \cite{gies} claim
that Al Khawaja {\it et al}. \cite{alkhawaja} argument that the
excitation of a single condensate atom is associated with an
energy of $-\mu_B$ includes on the mean-field energy of initial
and final states and neglects the other many-body effects on the
collision which presumably are included in the result
$g_{BB}=\langle 0|T_{2b}(-\mu_B)|0\rangle$. With this proviso we take
\begin{equation}    
g_{BB}=\frac{-4\pi\hbar^2}{m_B}\frac{1}{\ln{(\mu_B m_B
a_{BB}^2/4\hbar^2)}}
\label{int_2D-gbb}
\end{equation}
and similarly,
\begin{equation}    
g_{BF}=\frac{-2\pi\hbar^2}{m_{BF}}\frac{1}{\ln{((\mu_B+\mu_F) m_{BF}
a_{BF}^2/4\hbar^2]}}
 \label{int_2D-gbf}
\end{equation}
where the scattering lengths $a_{BB}=a_{BB}^{2D}$ and $a_{BF}=a_{BF}^{2D}$
are in principle 2D scattering lengths. Our choice for the 2D scattering 
lengths will be discussed in Sec. \ref{results}.
In the case of BF interaction strength we used the
reduced mass $m_{BF}$ and made the replacement $\mu\rightarrow
(\mu_B+\mu_F)/2$. Similar considerations to write down the
BF $T$-matrix were also made by Mur-Petit {\it et
al}. \cite{murpetit}.

In this regime the interaction parameters must be determined
self-consistently. The way the equations are written above suggests to 
solve for wave functions for given values of
$g_{BB}$ and $g_{BF}$, then calculate chemical potentials 
\begin{equation}
\mu_B=\frac{1}{N_B}\int\,d^2 r\left\{ \frac{\hbar^2}{2m_B}|\nabla\psi_B|^2+
\frac{1}{2}m_B\omega_B^2r^2|\psi_B|^2+g_{BB}|\psi_B|^4
+g_{BF}|\psi_F|^2|\psi_B|^2\right\}
\end{equation}
and
\begin{equation}
\mu_F=\frac{1}{N_F}\int\,d^2
r\left\{\frac{\hbar^2}{2m_F}\lambda_W|\nabla\psi_F|^2
+\frac{1}{2}m_F\omega_F^2r^2|\psi_F|^2+\frac{\hbar^2}{m_F}2\pi|\psi_F|^4
+g_{BF}|\psi_F|^2|\psi_B|^2\right\}
\end{equation}
and check whether the expressions for $g_{BB}$ and $g_{BF}$ are
satisfied. To follow common practice we start with initial chemical 
potentials, calculate $g$'s and then calculate chemical potentials using 
the obtained wave functions and require self-consistency. 
Note that in this regime the results do not depend on the value of
$\omega_{jz}$, i.e. on the value of the anisotropy parameter 
$\lambda=\omega_{Bz}/\omega_B$.

\subsection{Quasi-2D scattering}
When $a_z\gtrsim |a_{BB}|,|a_{BF}|$ collisions are bi-dimensional
but influenced by the
$z$-direction. In this regime, which is in between the previous cases, 
the 2D scattering
length can be expressed in terms of the 3D scattering length
\cite{Petrov2001}. Substituting
\begin{equation}    
a^{2D}_{ij} = 2\sqrt{2} \sqrt{\frac{\pi}{B}} a_z e^{-\sqrt{\pi/2}
\frac{a_z}{a^{3D}_{ij}}}
\label{2Dascatt}
\end{equation}  
in the coupling strength expressions for strictly
2D regime with $B\approx 0.915$, the coupling strengths now become
\cite{lee}
\begin{equation}    
g_{BB}=\frac
{\frac{2\sqrt{2\pi}\hbar^2}{m_B} \frac{a_{BB}}{a_z} }
{1 + \frac{1}{\sqrt{2\pi}}\frac{a_{BB}}{a_z} \ln ( B \hbar^2/2\pi \mu_B m_B
a_z^2)}
\label{int_Q2D-gbb}
\end{equation}  
and
\begin{equation}
g_{BF}=\frac
{\frac{\sqrt{2\pi}\hbar^2}{m_{BF}} \frac{a_{BF}}{a_z}}
{1+\frac{1}{\sqrt{2\pi}}\frac{a_{BF}}{a_z} \ln (B
\hbar^2/2\pi(\mu_B+\mu_F) m_{BF} a_z^2)}  .  
\label{int_Q2D-gbf}
\end{equation}

\begin{figure}[ht]
\begin{center}
\includegraphics[scale=1.4]{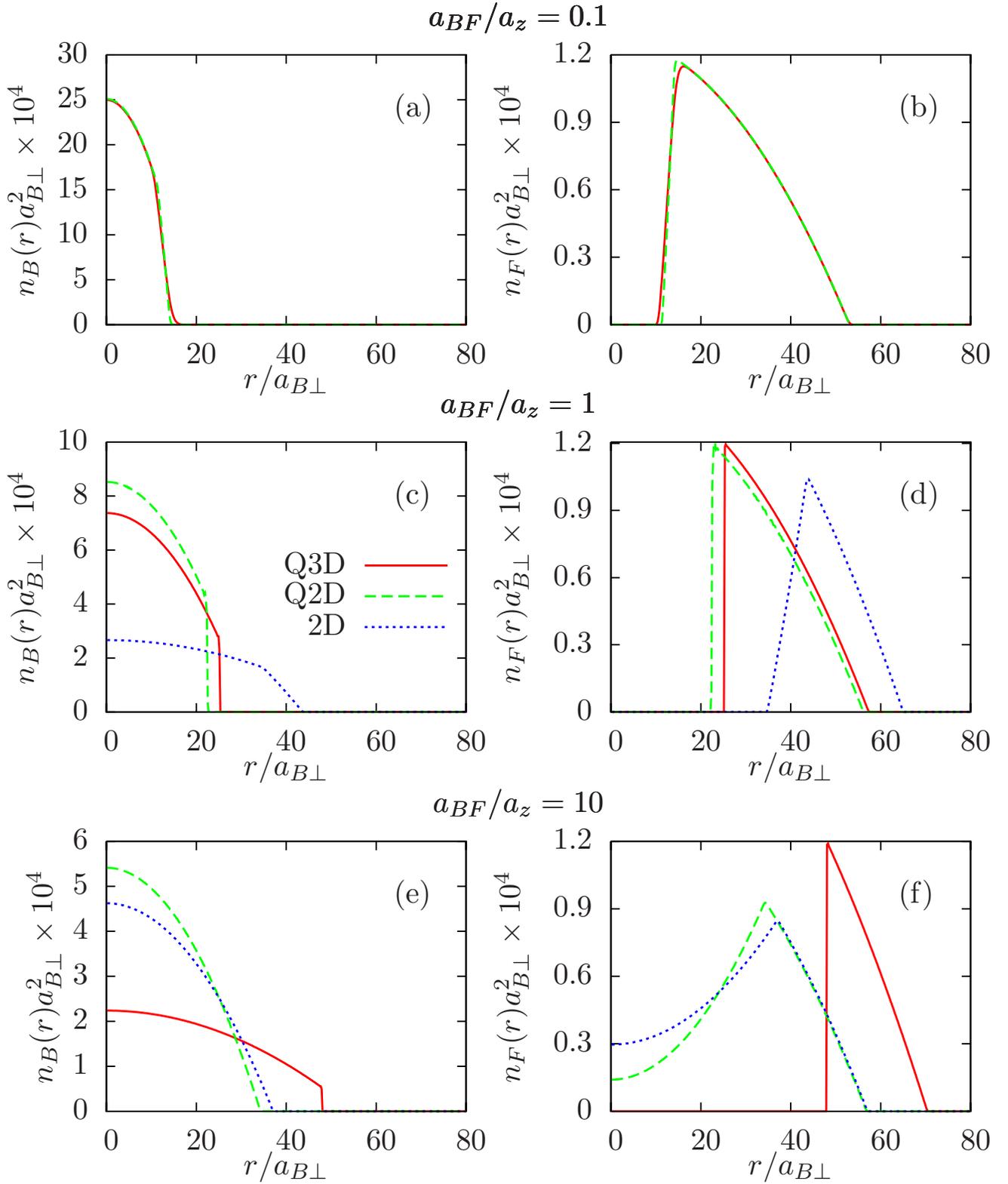}
\caption{(Color online)
Boson and fermion density profiles for $^6$Li-$^7$Li mixture
with $N_B=10^6$ and $N_F=5\times 10^5$,
radial trapping frequencies $\omega_B/2\pi=4000\,Hz$, 
$\omega_F/2\pi=3520\,Hz$ and scattering lengths
$a_{BB}=5.1\,a_0$, $a_{BF}=38\,a_0$ where
$a_0$ is the Bohr radius. The length unit is the radial harmonic
oscillator length for bosons $a_{B\perp}=\sqrt{\hbar/m_B\omega_{B}}$. The
density given is in units of $10^{-4} a_{B\perp}^{-2}$ and is
normalized to unity. The three regimes $a_{BF}/a_z=0.1,\,1,\,10$ 
correspond to values of the asymmetry parameter
$\lambda\approx 10^3,\,10^5$, and $10^7$,
respectively.
\label{fig1}}
\end{center}
\end{figure}

\section{Results and Discussion}\label{results}

We first consider a lithium mixture with particle numbers
$N_B=10^6$ and $N_F=5\times 10^5$, and radial trapping
frequencies $\omega_B/2\pi=4000$\,Hz and
$\omega_F/2\pi=3520$\,Hz. The BB and BF
scattering lengths are taken as $a_{BB}=5.1\,a_0$ and
$a_{BF}=38\,a_0$, respectively, in which $a_0$ is the Bohr
radius.

In Fig.\,\ref{fig1} we show the density distributions $n_B(r)$ and
$n_F(r)$ of bosonic and fermionic components in the three
scattering regimes: the quasi-3D, where the coupling is given by 
Eq.\,(\ref{int_3D}), the quasi-2D, where the coupling is given in 
Eqs.\,(\ref{int_Q2D-gbb}) and (\ref{int_Q2D-gbf}), and the strictly 2D, 
where we use the coupling given in Eqs.\,(\ref{int_2D-gbb}) and 
(\ref{int_2D-gbf}) and where
we set the bi-dimensional scattering lengths equal to $a^{2D}_{ij}$ 
(Eq.\,(\ref{2Dascatt})) evaluated
in the limit of vanishing $a_z/a^{3D}_{ij}$. This choice assures 
the strictly 2D model to be the limiting case of the Q2D, that depicts 
the crossover behavior.

When $a_{BF}/a_z=0.1$ (top panel) the mixture has 3D
character in terms of collisions even though the geometrical
confinement ($\lambda=10^3$) renders the system 2D
kinematically. The calculated chemical potentials
$\mu_B/\hbar\omega_{Bz}$ and $\mu_F/\hbar\omega_{Fz}$ being less
than unity also confirms that the system is geometrically 2D.
In this regime the density distributions for quasi-3D and
quasi-2D models look very similar. The boson and fermion 
components occupy the inner and outer
parts of the disk giving a segregated phase for the chosen
parameters. The 2D model is evidently inapplicable in this 
regime because $a_{BF}/a_z<1$.

In the middle panels of Fig.\,\ref{fig1} we show density profiles for 
the same mixture with $a_{BF}/a_z=1$ for an
anisotropy parameter $\lambda=10^5$. This corresponds to
a completely frozen motion in the $z$-direction and to
the crossover in the scattering properties from 3D to 2D. 
Figures \ref{fig1}(c) and \ref{fig1}(d) reveal that the
density profiles in the three models are very similar,
except for the fact that the 2D model predicts a larger spatial
extension of the density profiles. 

Finally, in the bottom panel of Fig.\,\ref{fig1} we consider
$a_{BF}/a_z=10$ with $\lambda=10^7$. $a_z$ being smaller than
in the previous case, the bi-dimensional scattering lengths are smaller
and both the 2D and Q2D models predict a mixed phase even in the 
center of the trap, while the Q3D curves still show phase separation.
For this anisotropy parameter, the
scattering events should be truly 2D and our corresponding
model should yield the most accurate
density profiles. Evidently the Q3D model is not yet valid, but we plot it
just to compare the predictions of the different models.


We now turn our attention to $^{40}$K-$^{87}$Rb mixture having 
an attractive BF scattering length. We consider a system with
particle numbers $N_B=10^6$ and $N_F=5\times 10^5$, and radial trapping
frequencies $\omega_B/2\pi=257$\,Hz and
$\omega_F/2\pi=378$\,Hz. The BB and BF
scattering lengths are taken as $a_{BB}=110\,a_0$ and
$a_{BF}=-284\,a_0$, respectively \cite{ospelkaus}. 
For attractive interactions, the effective 2D BF
scattering length is positive [see Eq.\,(\ref{2Dascatt})], 
namely the dimensional crossover 
induces effective repulsive interactions \cite{murpetit},
as already predicted
in a condensate with attractive boson-boson interaction \cite{Petrov2001}.
Thus, the strictly 2D couplings 
(Refs.\,\cite{lee,alkhawaja,gies,murpetit}) 
refer to hard-core collisions \cite{Schick1971}.

\begin{figure}
\begin{center}
\includegraphics[scale=1.4]{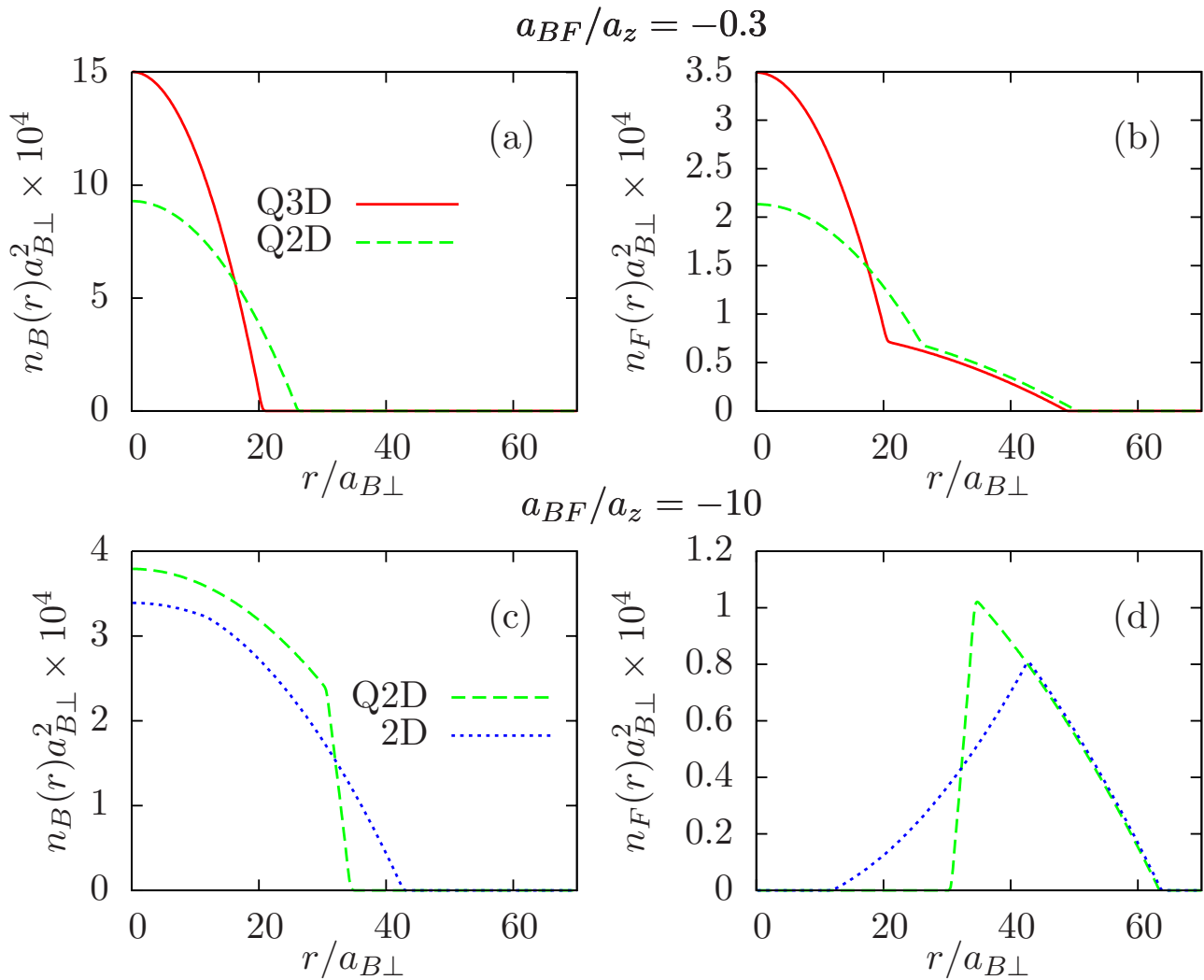}
\caption{(Color online)
Same as in Fig.\,\ref{fig1} for a $^{40}$K-$^{87}$Rb mixture with
$N_B=10^6, N_F=5\times 10^5$.
The values of $|a_{BF}|/a_z=0.3,\,10$ correspond to 
$\lambda\approx 2\times 10^2$ and $2\times 10^5$.
\label{fig2} }
\end{center}
\end{figure}

\begin{figure}
\begin{center}
\includegraphics[scale=1.4]{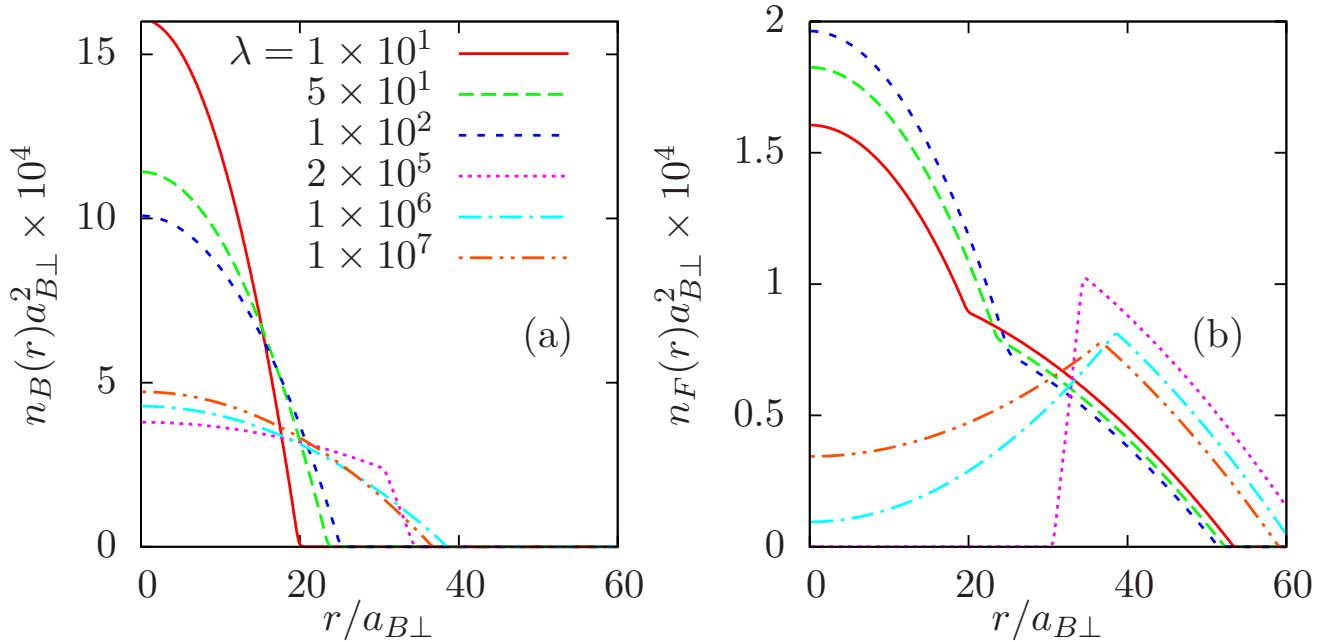}
\caption{(Color online)
Density profiles for 
the $^{40}$K-$^{87}$Rb mixture calculated with the Q2D model for 
various values of $\lambda$. (Same units as in Fig.\,\ref{fig1}.)
\label{fig3}
}
\end{center}
\end{figure}

\begin{figure}
\begin{center}
\includegraphics[scale=1.2]{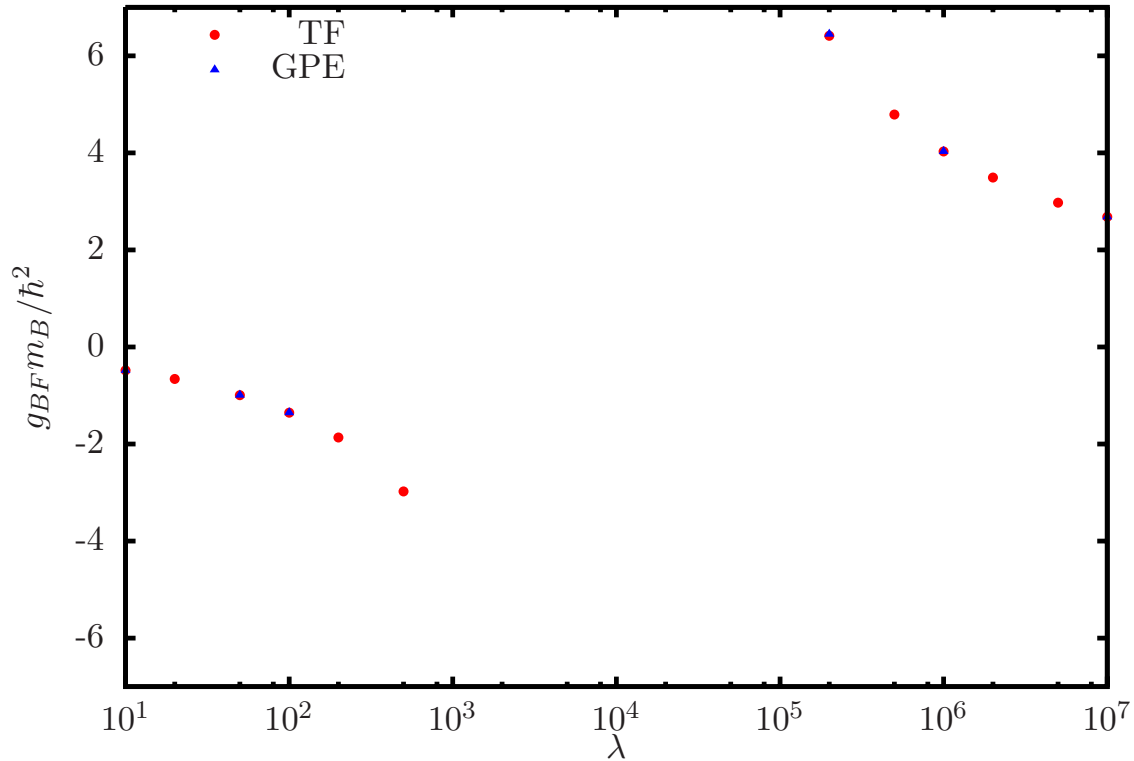}
\caption{(Color online)
Effective BF interaction strength (in dimensionless units) for the
$^{40}$K-$^{87}$Rb mixture within the Q2D model
as a function of the anisotropy parameter $\lambda$.
Dots refer to the numerical calculation performed in the 
Thomas-Fermi (TF) approximation, namely neglecting the Laplacian 
terms in Eqs.\,(\ref{dens1}) and (\ref{dens2}),
while triangles refer to the full solution of the same equations (GPE).
\label{fig4}}
\end{center}
\end{figure}

Figure \ref{fig2} illustrates the density profiles $n_B(r)$ and $n_F(r)$
in quasi-3D and 2D scattering regimes, characterized by
$a_{BF}/a_z=-0.3$ ($\lambda\approx 2\times 10^2$) and $a_{BF}/a_z=-10$
($\lambda\approx 2\times 10^5$), respectively. 
In the case $a_{BF}/a_z=-0.3$,
we observe that the density profiles are similar for
quasi-3D and quasi-2D models and show a bump in the center of
the fermionic density due to the attractions with the bosons.
For $a_{BF}/a_z=-10$, the Q2D model approaches the 2D one, 
the only difference being that the first model predicts complete 
spatial separation between the bosonic and the fermionic components, 
while the second predicts a residual mixed phase at the center of the trap.
The crossover between the two regimes is shown in Fig.\,\ref{fig3}.
For $\lambda<10^2$, the fermionic density is enhanced
at the center of the trap because of the presence of the bosons. 
In this regime the BF coupling term is negative, 
as shown in Fig.\,\ref{fig4}.  
At $\lambda=10^5$ the fermions are pushed out of the center 
of the trap because of the large
repulsive BF interaction (see Fig.\,\ref{fig4}). 
By increasing further and further
the anisotropy, the BF coupling is still positive 
but decreases and the two components are partially mixed. For
$10^2<\lambda<10^5$ no stable solutions are found. 
Thus, as shown in Fig.\,\ref{fig4}, the dimensional crossover plays
the role of a Feshbach resonance. Squeezing the trap one may 
naively expect
the gas just collapsing, but the crossover in the scattering geometry 
changes the nature of the instability from collapse to demixing, and a
further squeezing of the trap stabilizes the mixture.
All curves shown in this sections correspond to densities that fulfill
the diluteness conditions $n_Ba_{BB}^2\ll 1$ and $n_Fa_{BF}^2\ll 1$, even
at close to the resonance shown in Fig.\,\ref{fig4}.

\section{Summary}\label{summary}
In summary we have studied the equilibrium properties of a
boson-fermion mixture confined in a pancake-shaped trap, 
in the dimensional crossover from 3D to 2D.
The boson-boson and the boson-fermion couplings used are those derived
from the two-body $T$-matrix evaluated
(i) at zero energy in 3D, (ii) taking into account the discreteness
of the spectrum in the axial direction, in the crossover, (iii)
taking into account the many-body energy 
shift in the strictly 2D limit.
The density profiles and the couplings have been evaluated 
self-consistently using suitable modified coupled Gross-Pitaevskii 
equations for the bosonic and the fermionic wave functions.

For the case of a positive 3D boson-fermion scattering length, 
the dimensional crossover softens the repulsion, so that the 
components of a demixed boson-fermion mixture in 3D can mix in 
the 2D limit. For the case of a negative 3D boson-fermion scattering 
length, the dimensional crossover is more dramatic and plays 
the role of a Feshbach resonance. 
Our study shows that the squeezing of 
the pancake-shaped trap may drive a strong-attractive unstable
mixture towards a stable mixed mixture
passing through a demixed phase.
This numerical study may be reproduced in the actual experiments with BF 
mixtures. The goal being to reach a regime where the modulus 
of the scattering lengths is comparable or greater than the mixture axial 
size, one may exploit Feschbach resonances to increase the magnitude 
of the 3D scattering lengths, or one may engineer very flat
traps as already done in the context of experiments with a 
single BEC component.

\acknowledgments{
This work is supported by TUBITAK (No. 108T743), TUBA and
European Union 7th Framework project UNAM-REGPOT
(No. 203953).}

\end{document}